 \documentclass[final,5p,times]{elsarticle}
 \usepackage{graphics}
\usepackage{amssymb}
 \usepackage{lineno}

\newcommand{\degree}{\ensuremath{^\circ}}

\journal{NIM A  RICAP-2013}

\begin{document}

\begin{frontmatter}



\title{ Transient Point Source Analyses in the ANTARES Neutrino Telescope }


\author{ Agust\'in S\'anchez Losa on behalf of the ANTARES collaboration }

\address{ Agustin.Sanchez@ific.uv.es }

\begin{abstract}

  The ANTARES telescope, with a duty cycle close to unity and a full hemisphere of the sky at all the times visible, is well suited to detect neutrinos produced in astrophysical transient sources. Assuming a known neutrino production period, the background and the sensitivity can be drastically improved by selecting a narrow time window around it. GRBs, $\mu$-quasars and AGNs are particularly attractive potential neutrino point sources since neutrinos and gamma-rays may be produced in hadronic interactions with the surrounding medium as they are the most likely sources of the observed ultra high energy cosmic rays. A strong correlation between the gamma-ray and the neutrino fluxes is expected in this scenario.

  ANTARES data has been analysed in various transient source analyses with the goal of detecting cosmic neutrinos from GRBs, $\mu$-quasars and AGNs. The sensitivity of a standard time-integrated point source search can be improved by a factor 2-3 by looking for neutrinos only during the most probable emission time. This information can be provided by the different satellite telescope types on the X-rays and $\gamma$-rays wavelengths. The results of these different analyses will be presented.

\end{abstract}

\begin{keyword}
ANTARES \sep neutrino astronomy \sep transient analysis \sep micro-quasars \sep GRBs \sep flares \sep AGNs \sep blazars 

\end{keyword}

\end{frontmatter}

%  \linenumbers

\section{Introduction}
\label{Introduction}

  Several neutrino telescope experiments are currently doing point source analysis in the search for neutrino sources. Since the detected atmospheric neutrinos in these telescopes comprise an irreducible background, these searches have to be done by looking for an accumulation of events in a given source direction. This analysis can improve its performance substantially if the time window to search for neutrinos is limited to only the optimum time of emission of neutrinos at the source. Since the hadronic mechanism that can create a flux of neutrinos produces at the same time $\gamma$-rays, the time information of the neutrino emission can be inferred with the observed $\gamma$-ray flaring periods on those sources. This information can be provided by satellite telescopes, like FERMI, SWIFT or ROSSI.% and IACT (Imaging Air Cherenkov Telescopes) ground based (like HESS, MAGIC or VERITAS).

  Different variable sources can be observed in this way. Gam\-ma Ray Bursts (GRBs) are the most intense sources and the ones with the highest variability, which can reduce the neutrino time window to observe from a few seconds up to few days. Well identified sources which usually emit along the time $\gamma$-rays but with a variable flux can be studied. For galactic sources, the most promising ones are the so called $\mu$-quasars, compact objects with a companion star that can emit intense flares through irregular material accretion onto the compact companion. On the other hand, for extragalactic sources, Active Galactic Nuclei (AGNs) are the most promising candidates for such a variable flux emission, in particular the most intense and variable ones, the so called blazars (and specially the BL Lac's and the FRSQ objects). In both cases ($\mu$-quasars and AGNs) time variability goes from one day up to weeks.

  Here the three last analyses done by ANTARES in the search of neutrino emission in GRBs, $\mu$-quasars and AGNs are presented.

\section{The ANTARES neutrino telescope}
\label{ANTARES}

  The ANTARES neutrino telescope is placed at a depth of 2475~m on the sea bed of the Mediterranean Sea (42\degree48~N, 6\degree10~E). It is connected by a 42~km submarine cable to the shore of Toulon (France). 12 lines, separated by 60--70~m and vertically suspended by a buoy, are connected to this cable through a junction box. A single line is composed by 25 floors, except for line 12 which has only 20 floors, with a separation of 14.5~m between them. Each floor has a triplet of optical modules (OMs), each one housing a photomultiplier (PMT) facing 45\degree\ downwards. The full detector conforms a tri-dimensional array of 885 PMTs \cite{antares1} \cite{antares2}, completed in 2008 when the last line was connected and taking data with real-time processing. Around 7000 neutrinos have been detected since then, with a median angular resolution of 0.3-0.4\degree above $\sim$10~TeV and an effective area of $\sim$1~m$^{2}$ at 30~TeV. Three fourth parts of the sky are visible, including the Galactic Center and the most of the Galactic Plane.

  The neutrinos are detected via the Cherenkov light induced by the relativistic muons produced in the detector surroundings. The Cherenkov photons are detected in the array of PMTs where their arrival times and amplitudes are digitized (hits) \cite{Electronics} and sent to the shore station for muon reconstruction and physics analysis.

  From the reconstructed muons derive two sources of background: the atmospheric neutrinos produced in the cosmic rays (CRs) in the upper part of the Earth's atmosphere and the atmospheric muons from CRs that reach the detector from the above atmosphere. Although the latter can be suppressed by selecting only the up-going events in the detector. The atmospheric neutrinos are a dominant irreducible background. This implies that the search for a cosmic neutrino source with a neutrino telescope has to be done looking for accumulations of events in a certain direction and with a particular energy spectrum, harder than the one of atmospheric neutrinos. Only charged current interactions of neutrinos and antineutrinos has been considered.

\section{Transient sources analyses}
\label{Transient}

\begin{figure}[!t]
  %\vspace{5mm}
  \centering
  \includegraphics[width=7cm]{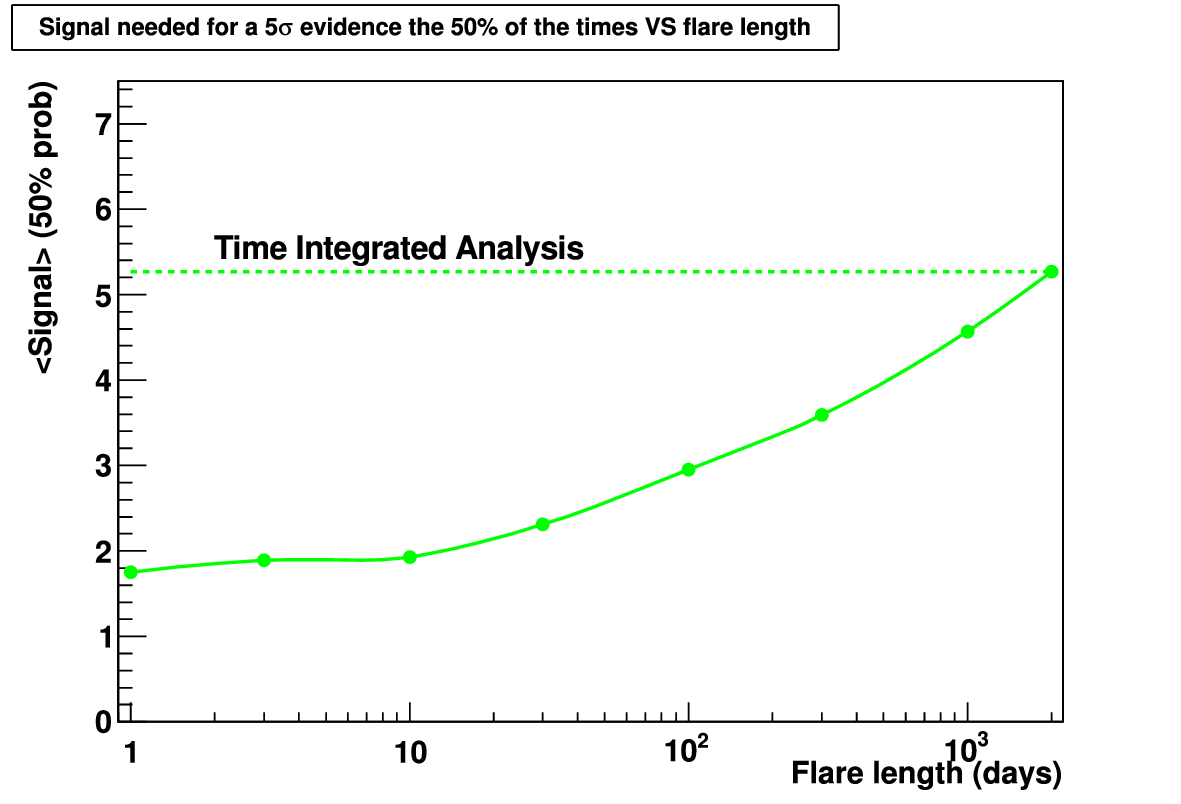}
  \caption{Average number of events required for a 5$\sigma$ discovery (50\% C.L.) of a source located at a declination of -30\degree\ as a function of the width of the flare period (solid line), a simple Heavyside function. This is compared to the number required for a time-integrated search (dashed line). The average typical length of the studied source flares in both AGN and $\mu$-quasars goes from 1 up to 100 days. GRBs last few hours. }
  \label{fig:discovery}
 \end{figure}

  In a point source search of neutrino sources, performance can be highly improved if the sought neutrino emission is constrained not only to a very particular source direction, but also to a very short time emission. That improvement could mean a factor 2 or 3 with respect to not using the time information at all, as it is shown in Fig.~\ref{fig:discovery}.

  This can be achieved in the frame of a multi-messenger study, where the expected neutrino time signal can be inferred from the $\gamma$-ray time information emission from the source. The motivation of this link lies on the Fermi acceleration mechanism, where in a very dense high energetic environment, in the presence of hadrons (protons mainly) and over a given energy threshold, neutral and charged pions start to be produced, decaying later in both photons and neutrinos. That is why, in the case of such hadronic acceleration scenario, a flux of neutrinos proportional to the flux of $\gamma$-rays is expected. Hence, when a burst of $\gamma$-rays is detected, an increase in the neutrino flux is expected, increasing the chances of detecting cosmic neutrinos during these $\gamma$-ray high emission periods. Of course, the neutrino emission enhancement will depend on the contribution of the hadronic acceleration mechanisms component over the leptonic one, this produce also $\gamma$-rays without neutrino emission, so constraints in the acceleration models can be deduced from the absence of neutrino signal.

  %The photon emission time information can be mainly provided by two kind of telescopes: satellites, like FERMI in $\gamma$-ray or SWIFT and ROSSI in X-rays, and IACT (Imaging Air Cherenkov Telescopes), in the range of high energy and ultra high energy $\gamma$-rays, like HESS, MAGIC or VERITAS. Different transient $\gamma$-ray emission candidates for neutrino production have been observed along the years. Probably the three most significant ones are the GRBs, the AGNs and the $\mu$-quasars, which analyses by ANTARES are presented right after.
  The photon emission time information can be provided by satellite telescopes, like FERMI in $\gamma$-ray or SWIFT and ROSSI in X-rays. Different transient high energy photon emission candidates for neutrino production have been observed along the years. Probably the three most significant ones are the GRBs, the AGNs and the $\mu$-quasars, which analyses by ANTARES are presented right after.

\subsection{GRBs}
\label{GRBs}

\begin{figure}[!t]
  %\vspace{5mm}
  \centering
  \includegraphics[width=7cm]{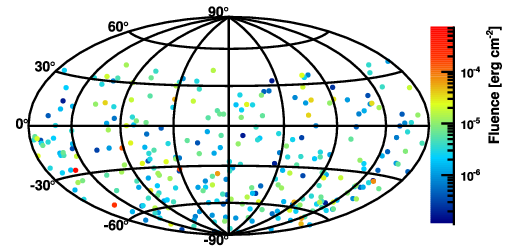}
  \caption{Sky distribution of the selected 296 gamma-ray bursts in equatorial coordinates. The photon fluence of each burst is indicated by the colors. }
  \label{fig:grb_skymap}
 \end{figure}

  The GRBs are the most energetic events known in the universe, their variability goes from a few seconds up to a few days. In this analysis only the so called long GRBs have been studied, since the physics behind the short GRBs is much less understood. It has been a stacked analysis with data from 2008 up to 2011, which comprises a total of 296 long GRBs, showed in Fig.~\ref{fig:grb_skymap}, during a total of 6.55 hours of live time, which time information has been provided by the FERMI, SWIFT and GCN (Gamma-ray Coordinates Network) alerts.

  An Extended Maximum Likelihood search has been done for the analysis, with the selection of events optimized for the highest discovery probability. The GRB simulations of the expected neutrino fluence have been done with two different model spectra: Guetta \cite{guetta} and NeuCosmA \cite{neucosma}. While the first one overestimates the pion production and hence the neutrino flux, the second is more conservative and it has been the one used for the optimization.

\begin{figure}[!t]
  %\vspace{5mm}
  \centering
  \includegraphics[width=7cm]{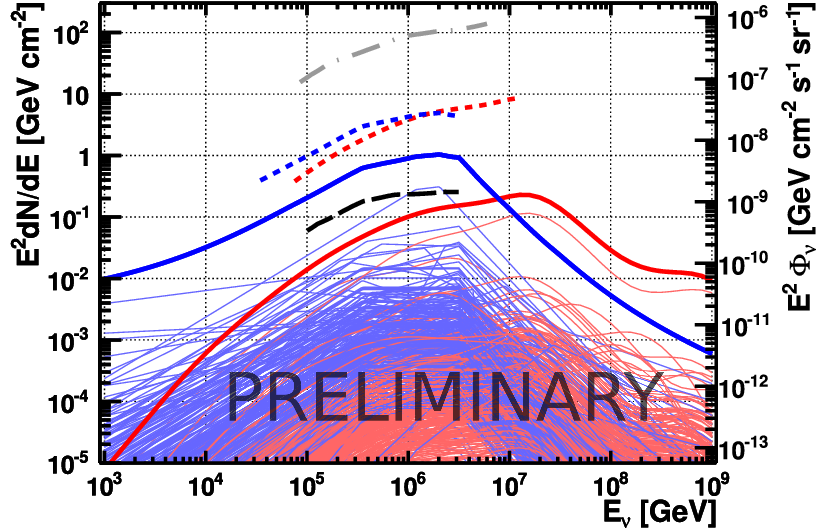}
  \caption{Upper limits (dashed lines) at a 90\% C.L. for Guetta (red) and NeuCosmA (blue) for the current GRB analysis, in comparison with the previous ANTARES limits \cite{oldgrb} (grey) and with IceCube IC40+59 limits \cite{icecubegrb} (black) for 215 GRBs. The remaining lines correspond to the simulated neutrino flux of each GRB with Guetta (red) and NeuCosmA (blue), being the thick lines the sum of all of them.}
  \label{fig:grb_results}
 \end{figure}

  No event has been found in the stacked GRB search windows, with an amount of expected events of 0.48 for Guetta and 0.061 for NeuCosmA spectra, meanwhile 0.05 events where expected from the background only hypothesis. Nevertheless, with respect to the previous ANTARES GRB analysis \cite{oldgrb} done for 40 GRBs, the upper limits have been improved as it is shown in Fig.~\ref{fig:grb_results}.

\subsection{$\mu$-quasars}
\label{mu-quasars}

  For this analysis six $\mu$-quasars, with outbursts in X-rays and $\gamma$-rays in the satellite data during the period 2007--2010, were studied: Circinus~X-1, GX~339-4, H~1743-322, IGRJ17091-3624, Cygnus~X-1 and Cygnus~X-3.

\begin{figure}[!t]
  %\vspace{5mm}
  \centering
  \includegraphics[width=8cm]{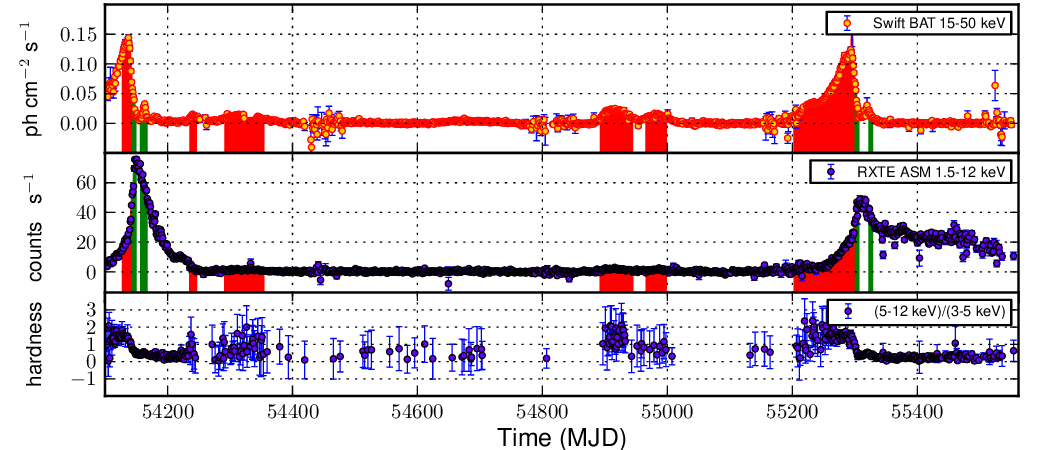}
  \caption{X-ray light curves of GX~339-4 between 2007 and 2010, where the identification of the hard states are shown in the red filled areas and the hard to soft transition in the green filled areas.}
  \label{fig:uq_lc}
 \end{figure}

  The neutrino search for the four black hole binaries has been split in two kinds: observing the source during the hard X-ray states, and during the transition from hard to soft emission states, both scenarios when the acceleration could be dominated by hadronic acceleration. Samples of those periods identifications are shown in Fig.~\ref{fig:uq_lc}. This X-ray emission information has been obtained from the SWIFT and ROSSI satellites, together with the one used for Circinus~X-1. For the $\gamma$-ray bursts of Cygnus~X-3, data from the Fermi LAT satellite has been used.

  The analysed period, by an unbinned method based on a likelihood ratio maximization, comprises the ANTARES data from 2007 up to 2010, a total of 813 days of live time. In this case no event has been found in time coincidence, but upper limits have been computed for the given sources, as it is shown in Table~\ref{tab:mu_results}.

\begin{table}
%\centering
\begin{center}
\begin{tabular}{|r|cc|}
  \hline
    Source & Livetime (days) & Fluence U.L.\\
  \hline
    Cir X-1 & 100.5 & 16.9 \\
    GX 339-4 (HS) & 147.0 & 10.9 \\
    GX 339-4 (TS) & 4.9 & 19.7 \\
    H1743-322 (HS) & 84.6 & 9.1 \\
    H1743-322 (TS) & 3.3 & 30.3 \\
    IGRJ17091-3624 & 8.5 & 21.3 \\
    Cyg X-1 (HS) & 182.8 & 14.1 \\
    Cyg X-1 (TS) & 18.5 & 6.0 \\
    Cyg X-3 & 16.6 & 5.7 \\
  \hline
\end{tabular} 
\end{center}
\caption{List of the $\mu$-quasars analysed by ANTARES. The Livetime is the effective time of observation of the source in days. Fluence is the Feldman-Cousins (90\% C.\,L.) upper limits on the neutrino fluence for a flux spectrum of $E_{\nu}^{—2}$ in GeV~cm$^{-2}$. }
\label{tab:mu_results}
\end{table}

\subsection{AGNs}
\label{AGNs}

  Other transient sources of interest analysed by ANTARES are the AGNs. Of particular interest are the ones that show the highest brightness and variability, conformed by the so called blazars subtypes BL Lac and FSRQ (Flat Spectrum Radio Qua\-sar). From the 1FGL (FERMI LAT 1-year Point Source Catalog) ten sources of the highest variability and luminosity that are visible by ANTARES have been selected. The neutrino arrival time information has been inferred from the $\gamma$-ray light curves provided by the FERMI LAT data. %The flaring periods of interest are selected, once the base line of the light curve is determined, by selecting

\begin{figure}[!t]
  \vspace{5mm}
  \centering
  \includegraphics[width=8.1cm]{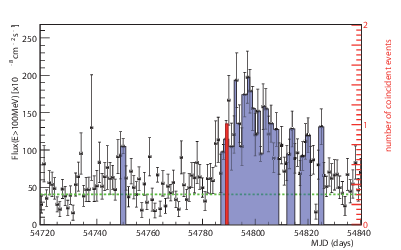}
  \caption{Gamma-ray light curve (black dots) of the blazar 3C279 measured by the LAT instrument on board the Fermi satellite above 100~MeV. Blue histogram: high state periods. Green dashed line: fit of a baseline. Red histogram: time of the ANTARES neutrino event in coincidence with 3C279. }
  \label{fig:pval}
\end{figure}

  The analysis comprises the period from September 6th up to December 31st of 2008, a total of 60.8 days of live time. %Only one neutrino event compatible in time and direction with one source was detected, specifically for the source 3C~279, at 0.56\degree from its position, and with a post-trial value of the 10\%. Also upper limits on neutrino fluxes were computed as is shown in Table~\ref{tab:agn_results}.
For it, an unbinned likelihood ratio maximization method has been used. The most significant source is 3C~279 with a pre-trial p-value of 1.03\%, it has been found one high-energy neutrino event during a large flare in November 2008 (Fig.~\ref{fig:pval}). This event (composed of 89 hits spread on 10 lines) has been reconstructed at 0.56\degree\ from the source location, with an estimated error of $\beta=0.3\degree$. The post-trial probability is computed taking into account the ten searches. The final probability of 10\% is compatible with a background fluctuation. Other source results are summarized in Table~\ref{tab:agn_results}.

\begin{table*}
%\centering
\begin{center}
\begin{tabular}{|l|ccccccc|}
  \hline
    Source & Class & Redshift & Visibility & Livetime (days) & $n_{5\sigma}$ & $n_{\mathrm{obs}}$ & Fluence U.L.$^{90\%C.L.}$ \\
  \hline
    PKS 0208-512  & FSRQ  & 1.003 & 1.00 & \ 8.8 & 4.5 & 0 & \ 2.8 \\
    AO 0235+164   & BLLac & 0.940 & 0.51 & 24.5  & 4.3 & 0 & 18.7  \\
    PKS 0454-234  & FSRQ  & 1.003 & 0.63 & \ 6.0 & 3.3 & 0 & \ 2.9 \\
    OJ 287        & BLLac & 0.306 & 0.39 & \ 4.3 & 3.9 & 0 & \ 3.4 \\
    WComae        & BLLac & 0.102 & 0.33 & \ 3.9 & 3.8 & 0 & \ 3.6 \\
    3C 273        & FSRQ  & 0.158 & 0.49 & \ 2.4 & 2.5 & 0 & \ 1.1 \\
    3C 279        & FSRQ  & 0.536 & 0.53 & 13.8  & 5.0 & 1 & \ 2.8 \\
    PKS 1510-089  & FSRQ  & 0.360 & 0.55 & \ 4.9 & 3.8 & 0 & \ 2.8 \\
    3C 454.3      & FSRQ  & 0.859 & 0.41 & 30.8  & 4.4 & 0 & 23.5  \\
    PKS 2155-304  & BLLac & 0.116 & 0.68 & \ 3.1 & 3.7 & 0 & \ 1.6 \\
  \hline
\end{tabular}
\end{center}
\caption{List of the bright, variable Fermi blazars selected for this analysis. $F_{300}$ is the gamma-ray flux above 300~MeV in $10^{-8}$~photons~cm$^{2}$s$^{-1}$. Live Time is the effective time of observation of the source in days. $n_{5\sigma}$ are the number of neutrino events needed to be detected in ANTARES with $5\sigma$, while $n_{\mathrm{obs}}$ is the number of events observed in the source direction and flare time window. Fluence is the upper limit (90\% C.\,L.) on the neutrino fluence in GeV~cm$^{-2}$, calculated according to the classical (frequentist) method for upper limits \cite{neyman}. }
\label{tab:agn_results}
\end{table*}

\section{Conclusions}
\label{Conclusions}

  ANTARES is suited for perform several multi-messenger analyses. Here have been presented the ones based on the link motivated by the expected correlation of neutrino and $\gamma$-ray emission in hadronic scenarios. The time information, provided by different $\gamma$-ray telescopes, reduces significantly the background in the cosmic neutrino search. This circumstance can be exploited in three source types of interest due to their variable $\gamma$-ray emission: GRBs, $\mu$-quasars and AGNs. ANTARES analyses on these sources have been presented.

  GRBs upper limits have been improved with respect to the previous analysis, while the first results for $\mu$-quasar analysis with ANTARES has been presented. The AGN analysis has the most significant observation, with a neutrino event, for 3C279 with a p-value of about 10\% after trials. In all the cases upper limits have been also provided. For the near future, AGN analysis is being currently updated with the ANTARES data up to 2012, including some improvements in the flaring period selection method, IACT (Imaging Air Cherenkov Telescopes, like HESS, MAGIC or VERITAS) detected flares, various energy spectrum analysed beside the $E_{\nu}^{—2}$ and a new parameter that allows a delay between the neutrino and the $\gamma$-ray signal, of special relevance for very short flares.

\section{Acknowledgments}
\label{acknowledgments}
  %Special thanks to Tonino, for saving us of getting totally dumped, on an expectedly strong storm afternoon, by taxing us back to the hotel and let us him umbrella. Also I'm deeply grateful for the efforts of Giulia and Chiara in order to helping me to print out and rescan once signed the contract of my Marseille apartment rent, they saved my file. Also to Juande for his efforts for correct my terrible English and in gave me some publicity on talks, but specially because of his closeness and orienteering in this business, specially needed in my case and hugely welcome on those days, it's obviously good to have such a referee. And probably to Juanjo also, such a huge patience with someone that is doing now the military service equivalent, he will manage to make me learn how to think carefully my words before answer a work question...
We gratefully acknowledge the financial support of the Spa\-nish Ministerio de Ciencia e Innovaci\'on (MICINN), grants FPA\-2009-13983-C02-01, FPA2012-37528-C02-01, ACI2009-1020, Consolider MultiDark CSD2009-00064 and of the Generalitat Valenciana, Prometeo/2009/026.



%% References


\bibliographystyle{elsarticle-num}
% \bibliography{<your-bib-database>}



\end{document}